# The Human Cell Atlas & Equity: Lessons Learned

**Standfirst: The Human Cell Atlas has been undergoing a massive effort to support global scientific equity. The co-leaders of its Equity Working Group share some lessons learned in the process.**

Partha P. Majumder Ph.D[1,2], Musa M. Mhlanga Ph.D[3,4,5], and Alex K. Shalek Ph.D[6,7,8]

[1]Indian Statistical Institute, Kolkata, India
[2]National Institute of Biomedical Genomics, Kalyani, India
[3]Radboud Institute for Molecular Life Sciences (RIMLS), Radboud University Medical Center, Nijmegen, The Netherlands.
[4]Epigenomics & Single Cell Biophysics Group, Department of Cell Biology, Radboud University, Nijmegen, The Netherlands
[5]Department of Human Genetics, Radboud University Medical Center, Nijmegen, The Netherlands
[6]Institute for Medical Engineering & Science, Department of Chemistry, and Koch Institute for Integrative Cancer Research, Massachusetts Institute of Technology, Cambridge, MA
[7]Ragon Institute of MGH, MIT, and Harvard, Cambridge, MA
[8]Broad Institute of MIT and Harvard, Cambridge, MA

Correspondence: Partha P. Majumder, ppm1@nibmg.ac.in; Musa M. Mhlanga, musahca@mhlangalab.org, Alex K. Shalek, shalek@mit.edu

Recent political and social events, primarily those originating in the United States, have triggered an intense desire for equity in all facets of the human experience. More specifically, actions engendered by the Black Lives Matter movement and others have led to the scrutinization of equity across a wide range of fields, from politics and business to academia and scientific research. In science, in particular, several major journals have published opinion pieces and editorials seeking greater equity or relating the "non-white" experience[1–3]. Many of their readers have been stunned by the revelations. Indeed, the scientific community is only now coming to terms with an unsettling and uncomfortable truth: structural exclusion of non-white people permeates all levels of the scientific enterprise. That said, with awareness comes opportunity. New frameworks for describing and addressing these issues have recently emerged[4,5], creating a structure within which groups can each consider how to best internalize and embody the lessons in their own scientific initiatives.

In the Human Cell Atlas (HCA) consortium, equity has been a point of emphasis from inception in 2016 for one simple reason: the HCA's success depends upon it. Fundamentally, the HCA is meant to be a foundational resource, inclusive of the many cell types and states found in healthy individuals across the globe[6]. That resource can then be used to address a wide range of scientific questions and, in the future, to facilitate a better understanding of disease. This mission demands, explicitly, the inclusion of representation along axes of gender, age, ethnicity, environment, socioeconomic status, and, in some cases, disease susceptibility in its biospecimens. Moreover, it requires broad **participation** to ensure comprehensive coverage, identify barriers to success, and support continuity, and necessitates reciprocal, balanced **benefit from the methods, data, and results** to ensure global engagement.

To this end, the HCA has set ambitious and dynamic equity goals for itself. Below, we describe key lessons learned through equity activities to date, as well as our future plans.

1. Trust must be given and earned
*For the HCA to be representative of the people, it must be erected by the people.* At its core, equity is about partnerships. Ensuring comprehensive representation in the HCA necessitates global collaboration to facilitate coverage, address potential regional barriers to success and, given the scope and timeline of the effort, support continuity and sustain institutional capacity. Fostering scientific participation among such a diverse and inclusive community can also help spur innovation, accelerating and strengthening efforts underway.

Critically, partnerships are built on mutual trust. Given power imbalances and a history of structural exclusion, those in positions of privilege must recognize the potential barriers associated with the development of that trust and, themselves, actively trust early and often, while working to earn the same from those who have been systematically marginalized.

*And it must be for the people.* Ensuring that scientific partnerships result in equitable—not equivalent—benefit to those involved can help to earn this trust, even in the face of historical inequities. To inspire individuals to donate biospecimens to, and partake in, efforts like the HCA, these activities must offer reciprocal, balanced benefit from the methods, data and results they generate to all. For each initiative, this means making its methods, data, and results as openly accessible as possible (within the context of all applicable ethical and legal standards) to those who have been systematically excluded, while simultaneously recognizing and protecting their work and interests.

Moreover, it means carefully considering local priorities. In many regions, for example, a healthy human cell atlas is only easily justified as a reference against which to understand human maladies because of the priorities of local funders. This has become all the more acute in light of the COVID-19 pandemic. For regions of the world that bear a disproportionate burden of disease, advancing the HCA while contemporaneously using its tools to elucidate the cellular basis of various ailments toward realizing new therapeutics and prophylactics will be essential to ensuring equivalent benefit.

More broadly, a deep understanding of local aspirations, opportunities, and challenges is best ingrained through active outreach and inquiry. For the HCA, this means not only a combination of hands-on training programs, workshops, conferences, and dissemination initiatives among different sectors of society in diverse geographical regions but also active questioning and listening.

## 2. Ask questions and listen carefully
To understand and support local needs, it is critical to be inquisitive and listen thoughtfully. For example, the goals of the HCA's initial Equity Strategy Meeting and first official Equity meeting were, ostensibly, to help introduce the HCA, and the transformative opportunities it affords, to the global health community, and to further consider fundamental compositional, organizational, and educational goals that could help the HCA to achieve equity. Yet, unlike most HCA events to date, the majority of the attendees were unfamiliar with the HCA (and many of its core technologies). These meetings therefore provided a critical chance to: 1. more comprehensively understand the goals and motivations of those involved in human biological research (in one capacity or another) around the world; 2. solicit engagement in HCA activities; and, 3. understand potential barriers (political, social, or otherwise) to the HCA's realization around the world. What we quickly learned by asking questions and listening was that every country and region has its own unique set of opportunities and challenges, and that our solutions would therefore need to be customized to each situation to help ensure that our impact matched our intentions.

Although these and related events impressed upon us the enormity of the challenge, they also highlighted widespread enthusiasm for participating in the HCA and bringing its relevant benefits to different national contexts. This, in turn, begged the question: what are the barriers that prevent such a palpable desire to equitably engage from materializing?

**3. Inequity can be driven by lack of knowledge.**

Often, information, approaches, and initiatives we assume to be common knowledge are, in fact, not. The HCA, illustratively, is anchored on concepts and methods that are at the frontier of basic science, only having been realized in the past five years or so. The specific background required for a deep understanding of these, as well as the initial scientific results being derived by the laboratories engaged in the HCA (which are mostly located in the high-income countries (HICs)/the global north), may not be readily available in low- and middle-income countries (LMICs)/the global south. As a result, in spite of keen aspirations among scientists working in LMIC institutions to participate and contribute to the goals of HCA, their restricted knowledge base has proved to be a barrier. This was clearly articulated during discussions at the HCA Equity meeting in Addis Ababa and reinforced in smaller group conversations, and must be overcome to promote equitable participation in the HCA.

To begin to examine how to help build this knowledgebase, the first HCA Equity Roadshow was held at the University of São Paulo (USP) in Brazil in September, 2019 to introduce members of the Brazilian scientific community to the HCA and explore potential intersections between the HCA and local research priorities. This meeting initiated conversations with both USP and local funding agencies to help support efforts in São Paulo, Brazil, and, more broadly, Latin America, catalyzed the creation of a local HCA community and HCA South America (dramatically increasing engagement from Latin America), and directly lead to USP's hosting the first HCA general meeting in Latin America in September, 2020.

Overall, the engagement in Brazil highlighted one initial, highly impactful action the HCA could pursue: by focusing on roadshows and workshops dedicated to outreach and training in geographic areas underrepresented in the HCA, the HCA could broaden its exposure and empower local students, postdoctoral fellows and principal investigators. Whether hands-on or remote, these efforts could help impart critical experimental and computational know-how. During these events, it is important to teach but also to *listen and learn* to help forge lasting partnerships, nucleate a sense of community, and help empower local scientists to take action and ownership in the realization of the HCA, its activities and policies. Moreover, these events require a mixture of advanced and lay education. Without a general awareness of the HCA's goals and possible benefits, the collection of the biospecimens required to make the atlas a reality, even if done under "informed" consent, will be an imposition and thus undermine equity.

To maximize impact and promote self-sustainability, these sessions should be designed to emphasize training-the-trainer. Additionally, support and facilitation of bidirectional travel for short- and long- term visits are critical to ensure the dissemination of HCA techniques from expert labs, as well as to seed and guide execution of pilot projects. Collectively, these actions can help ensure that underrepresented constituencies, at least in the HCA, are in a position to create locally relevant data sets. But, equity requires more than information – it requires physical capacity too.

**4. Inequity can be driven by lack of infrastructure.**
In many parts of the world, a lack of scientific infrastructure results in the structural exclusion of specific groups from both the HCA and genomics more broadly. Evidently, the HCA thrives in environments where human capacity and skill sets are sufficiently developed. This, in turn, requires scientific infrastructure capable of supporting single-cell work. One straightforward strategy to facilitate local empowerment is to identify, partner with, and fortify existing local infrastructure in regions underrepresented in the HCA, such as the West African Centre for Cell Biology of Infectious Pathogens (WACCBIP) in Ghana. Such centers can become hubs for training that can host future educational workshops and also provide a base for local, and potentially regional, processing and analysis of biospecimens.

A related part of the problem is timely access to affordable reagents. To enable wider participation, pricing deals must be negotiated with manufacturers to help reduce barriers to entry.

**5. Inequity can be driven by lack of real or perceived capacity.**
In some places, infrastructure exists; in others, it has yet to be developed creating a more significant roadblock to engagement. Beyond considering physical obstacles, our Equity meetings to date have underlined the importance of training local talent to utilize such infrastructure to empower deeper scientific engagement. In some underrepresented areas, this can require the development of mechanisms to seed the execution of pilot projects to help establish local scientific self-confidence. Often a lack of educational, physical, and/or financial infrastructure leads directly to diminished scientific participation and contribution, even in situations where capacity exists or could exist. The positive results of local pilot activities can, moreover, demonstrate to those in LMICs/the global south that they can be equal participants. Encouragement and support from those in HICs are essential to ensure that these activities help establish mutual trust and inspire deeper, more meaningful and balanced partnerships.

**6. Identify and articulate how equitable partnerships, on multiple levels, can lead to long-term empowerment, mutual benefit and equal participation.**
As flag bearers for the HCA and other efforts arise locally and around the world, we must be their champions and remain ever mindful of the barriers that may stand in their way, from the availability of affordable and timely reagents and equipment, to access to scientific literature, to uncertain career prospects and exclusionary practices, such as the failure to properly credit or attribute work, especially for team science.

Moreover, we must never fail to monitor our own progress with specificity. The HCA's ultimate goal is to generate an atlas of the human body that is representative of humankind. Therefore, we must quantify and track how well human diversity – ethnic, educational, socio-economic, age and gender, and geographical – is represented in the biospecimens that comprise the HCA. In addition to monitoring biospecimen submissions and publications, we must equally chart and continually assess the diversity of scientists who are participating in and contributing to HCA research, collaboratively and independently. Further, we must examine outward facing activities, such as the number of outreach and technical training programs held annually and those served by them, as well as those critical to the HCA's scientific success, such as the allocation of funding and experimental and computational infrastructure improvements. Finally, the HCA must also consider the connectivity among its researchers -- the number of collaborative edges between countries on publications, presentations, and grants. While these metrics might not be flattering initially, our hope is that they will inspire all those involved in scientific research, now and in the future, to work toward their active improvement.

At the end of the day, the rapid pace of innovation in the HCA and beyond is widening the scientific knowledge gap between those in HICs/the global north and LMICs/the global south. This only further entrenches systematic exclusion in the sciences. Creating a scientific culture that dissolves these systemic inequities by actively seeking to increase representation, participation, and benefit from the data and methods will be challenging, but it has boundless potential to transform and accelerate scientific understanding. The time to act is now.

The authors declare no competing interests.

**Acknowledgements**: The authors would like to thank all of the members of HCA, its Organizing Committee, and its Equity and Ethics Working Groups; all the individuals who participated in the HCA Equity Strategy Meeting at the Wellcome Trust in London, England, the HCA Brazilian Roadshop, and the first official HCA Equity meeting in Addis Ababa, Ethiopia; the Bill & Melinda Gates Foundation, The Wellcome Trust, The Ragon Institute, The Broad Institute, and The Chan Zuckerberg Initiative for their support; Jonah Cool and Mark Gilligan; Finally, we are grateful to all of the individuals and scientists who have participated and will participate in the HCA initiative and its quest for equity through donating their biospecimens, time, and/or efforts.